\PassOptionsToPackage{table}{xcolor}
\documentclass[sigconf,nonacm,natbib=true]{acmart}
\AtBeginDocument{%
  }

\usepackage{booktabs}
\usepackage{enumitem}
\usepackage{threeparttable}
\usepackage{tabularx}
\newcolumntype{C}{>{\centering\arraybackslash}X}
\newcolumntype{R}{>{\raggedleft\arraybackslash}X}
\newcolumntype{L}{>{\raggedright\arraybackslash}X}
\usepackage{graphicx}
\usepackage{xcolor}

\makeatletter
\newcommand{\acmrightssize}{\fontsize{8}{9.5}\selectfont}
\newcommand{\firstpagerights}[1]{%
  \begingroup
    \renewcommand\thefootnote{}%
    \footnotetext{%
      \acmrightssize
      \raggedright
      \setlength{\parskip}{0pt}%
      \setlength{\parindent}{0pt}%
      #1%
    }%
    \addtocounter{footnote}{0}%
  \endgroup
}
\makeatother

\begin{document}

\title[AutoSim: A Self-Improving User Simulator for Conversational Search Evaluation]{\textsc{AutoSim}: A Self-Improving User Simulator\\for Conversational Search Evaluation}

\author{Saber Zerhoudi}
\orcid{0000-0003-2259-0462}
\affiliation{%
  \institution{University of Passau}
  \city{Passau}
  \country{Germany}
}
\email{szerhoudi@acm.org}

\author{Jelena Mitrovic}
\orcid{0000-0003-3220-8749}
\affiliation{%
  \institution{University of Passau}
  \city{Passau}
  \country{Germany}
}
\email{jelena.mitrovic@uni-passau.de}

\author{Michael Granitzer}
\orcid{0000-0003-3566-5507}
\affiliation{%
  \institution{Interdisciplinary Transformation University Austria}
  \city{Linz}
  \country{Austria}
}
\affiliation{%
  \institution{University of Passau}
  \city{Passau}
  \country{Germany}
}
\email{michael.granitzer@uni-passau.de}

\begin{abstract}
User simulators promise to evaluate conversational search systems at a scale human studies cannot reach, but only if their behavior is realistic enough to trust. We study \textsc{AutoSim}, a conversational-search user simulator whose relevance judgments rank a controlled pool of retrieval systems in agreement with NIST across three TREC iKAT years (Kendall's $\tau$ of $0.66$, $0.91$, and $0.91$; all $p{\le}0.002$), an agreement that
survives removing its persona. \textsc{AutoSim} self-improves against a held-out signal from real users, and we find that fine-tuning its weights makes it behave like real users, which in turn gives a more accurate estimate of how well a system serves them. 
Reaching that behavior takes more than prompting: when discriminating between real and simulated user sessions, no prompt we tried lowers a held-out real-versus-simulated classifier below AUC $0.85$, whereas fine-tuning the weights drives it to $0.45$, where neither the classifier nor a human reading full 
dialogues can tell it from a real user. That added realism is what makes the evaluation more accurate: where real users retrieve little, a prompted simulator over-estimates retrieval effectiveness by $+0.136$ nDCG and returns a relevant passage on $57\%$ of the turns where real users find none, enough to change a deployment decision. Only the weight-level simulator escapes this bias. We release the framework and code\footnote{Code, evaluation, and models: \url{https://github.com/searchsim-org/sigir-ap26-autosim}\label{fn:repo}.}.
\end{abstract}

\begin{CCSXML}
<ccs2012>
   <concept>
       <concept_id>10002951.10003317.10003331</concept_id>
       <concept_desc>Information systems~Users and interactive retrieval</concept_desc>
       <concept_significance>500</concept_significance>
       </concept>
   <concept>
       <concept_id>10002951.10003317.10003359</concept_id>
       <concept_desc>Information systems~Evaluation of retrieval results</concept_desc>
       <concept_significance>500</concept_significance>
       </concept>
 </ccs2012>
\end{CCSXML}

\ccsdesc[500]{Information systems~Users and interactive retrieval}
\ccsdesc[500]{Information systems~Evaluation of retrieval results}

\keywords{User simulation, Conversational search, Evaluation, Self-improving agents}

% \copyrightyear{2026}
% \acmYear{2026}
% \setcopyright{cc}
% \setcctype{by}
% \acmConference[SIGIR-AP 2026]{Proceedings of the 2026 Annual International ACM SIGIR Conference on Research and Development in Information Retrieval in the Asia Pacific Region}{December 13--15, 2026}{Singapore, Singapore}
% \acmBooktitle{Proceedings of the 2026 Annual International ACM SIGIR Conference on Research and Development in Information Retrieval in the Asia Pacific Region (SIGIR-AP 2026), December 13--15, 2026, Singapore, Singapore}
% \acmDOI{10.1145/3848000.3850196}
% \acmISBN{979-8-4007-3033-7/2026/12}

\maketitle

\firstpagerights{%
  \textcopyright{} 2026, held by the owner/author(s); published under a Creative Commons Attribution 4.0 International (CC BY 4.0) License.\\
  This is the author's version of the work. The Version of Record was published in the \emph{Proceedings of the 2026 Annual International ACM SIGIR Conference on Research and Development in Information Retrieval in the Asia Pacific Region (SIGIR-AP 2026), December 13--15, 2026, Singapore, Singapore}, and is available online at: \url{https://doi.org/10.1145/3848000.3850196}.
}

\section{Introduction}
\begin{figure}[t]\centering
\vspace{2em}
\includegraphics[width=0.93\columnwidth]{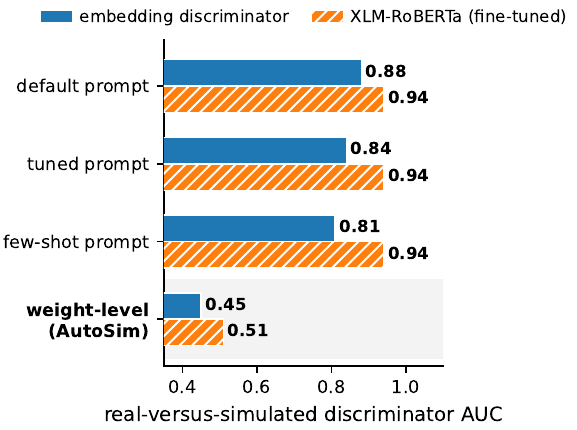}
\caption{Real-vs-simulated discriminator AUC for four simulator configurations on held-out iKAT: a default prompt, a tuned prompt, few-shot prompting, and a weight-level fine-tune (\textsc{AutoSim}), each scored by an embedding discriminator and a fine-tuned XLM-RoBERTa discriminator. An AUC of $0.5$ means real and simulated utterances are indistinguishable; higher is more separable.}\label{fig:substrate}
\end{figure}

User simulation is a standard way to evaluate interactive and conversational IR systems without running a human study for every design choice. A simulator stands in for a population of users, so retrieval and response strategies can be compared reproducibly when tuning a ranker or a clarifying-question policy.

Because a simulator is fixed at design time as a set of personas, prompts, and decision rules, its behavior drifts from that of real users as the systems it evaluates change. Worse, even where it does not drift, a simulator is a documented unreliable proxy: changing only the underlying user model swings the evaluated agent's measured task-success rate by about nine percentage points, calibration against humans is poor, and general-purpose LLM simulators systematically over-estimate success~\cite{Seshadri:2026:arXiv,Zhou:2026:arXiv}. The problem runs deeper than tuning: a more realistic simulator is not necessarily a more accurate evaluator~\cite{Balog:2025:SIGIR}.

This gap shows up directly in the utterances: asked to play a search user, a strong LLM writes long, polite, fully formed requests, whereas real users are brief and lean on the prior conversation for what they leave unsaid. Measuring it on TREC iKAT (\S\ref{sec:diagnosis}) locates the gap in what the simulated user says and asks for rather than in how the simulator judges relevance, and shows that editing the generation prompt does not close it, which is what tells us where self-improvement has to act.

\begin{figure*}[t]\centering
\centering
\includegraphics[width=0.85\textwidth]{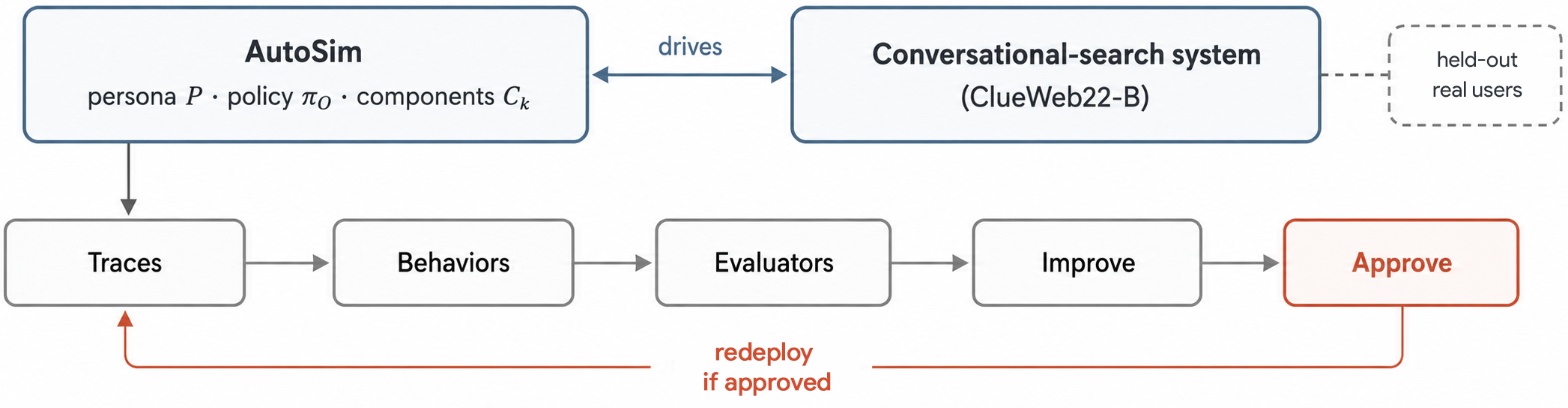}
\caption{\textsc{AutoSim} drives a real conversational-search system. The loop records the simulator's traces and proposes edits, accepting an edit only if it does not regress on a held-out split. The loop can edit two components: the relevance-judging instruction and the utterance generator.}\label{fig:arch}
\end{figure*}

Two lines of work relate to this problem, and they have developed independently. Trained user models are more realistic than prompted ones~\cite{Naous:2025:arXiv,Zhu:2025:arXiv}, and self-improving agents raise their own behavior by editing themselves, from prompts to weights~\cite{Agrawal:2025:arXiv,Zweiger:2025:arXiv}; neither asks what a more realistic simulator does to evaluation validity. We ask whether a simulator can close its own gap to real users, whether that closing has to happen at the prompt or the weights, and whether closing it makes the simulator a more accurate evaluator. Therefore, we study \textsc{AutoSim}, a hybrid orchestration-based simulator that self-improves in a loop checked against a held-out signal from real users (Fig.~\ref{fig:arch}), and we vary what the loop edits: the generation prompt, or the model weights. The two behave differently (Fig.~\ref{fig:substrate}). A prompt-level loop does not reduce the behavioral gap below the level a single hand-written prompt already reaches, whereas a weight-level loop reduces it until the discriminator can no longer separate the two. 
% Only the weight-level change reduces the simulator's bias as an evaluator.

\noindent We make three contributions, each backed by its own experiments.

\noindent\textbf{(1) A validated evaluator} (\S\ref{sec:validity},~\ref{sec:rq1}). \textsc{AutoSim} ranks real conversational-search systems in agreement with NIST across three iKAT Track-years ($\tau{=}0.66$ on 2023\cite{Aliannejadi:2024:TREC}; $\tau{=}0.91$ on 2024 and on the multi-session 2025 \cite{Aliannejadi:2025:TREC}; all $p{\le}0.002$), with a re-scoring standard deviation of $0.01$, with two real participant runs added ($\tau{=}0.65$), under a second assessor family, and on a second corpus (CAsT 2020\cite{Dalton:2020:TREC}, $\tau{=}0.56$, rising to $0.58$ when a neural reranking tier is added to the pool). The agreement comes from the relevance-assessment step, which is why it persists when the persona is removed.

\noindent\textbf{(2) A limit that depends on where self-improvement acts} (\S\ref{sec:substrate}). Trained user models are known to be more realistic than prompted ones~\cite{Naous:2025:arXiv,Zhu:2025:arXiv}; we measure where the prompt reaches its limit. Prompt-level self-improvement does not lower the real-versus-simulated discriminator below an AUC near $0.85$, and a single hand-written prompt matches an iterative loop and few-shot prompting. Weight-level self-improvement lowers the AUC to $0.45$ with human-level lexical diversity; the drop replicates on a second model family (Mistral-7B) for communication style, and the match extends to four behavioral measures on the primary model.

\noindent\textbf{(3) Realism changes the evaluation} (\S\ref{sec:bias}). Where real users' own queries retrieve few relevant passages, a prompt-level simulator over-estimates retrieval effectiveness by $+0.136$ nDCG. Matching length is not enough: a terse prompt still over-estimates by $+0.087$, and only the weight-level simulator is not detectably biased (point estimate $+0.014$, confidence interval including zero). The over-estimate hides retrieval failures, can change a deployment decision at a modest quality threshold, and does not appear in the summary metrics the field usually reports. We release the framework (LLMs with SPLADE, BM25, and a cross-encoder over ClueWeb22-B and NIST qrels), the evaluation protocol, and a blind human-discrimination study with its kit~\footref{fn:repo}.

Section \S\ref{sec:related} positions the work; \S\ref{sec:diagnosis} measures where the gap lives; \S\ref{sec:autosim} describes the simulator, the two components the loop can edit, and the validity protocol; \S\ref{sec:eval} answers the three questions in order.

\section{Related Work}\label{sec:related}
Self-evolving agents improve from their own experience by editing different parts of the system: prompts via reflective evolution~\cite{Agrawal:2025:arXiv}; memory via experience distillation~\cite{Wu:2025:arXiv}; weights via self-edits~\cite{Zweiger:2025:arXiv} or self-play~\cite{Huang:2025:arXiv,Zhao:2025:arXiv}; and the reward model itself via self-rewarding or self-taught evaluation~\cite{Yuan:2024:arXiv,Wang:2024:arXiv}. A recurring lesson is that the strongest gains come when the outer-loop signal is a real held-out performance signal, which we build in through a held-out acceptance check. These works establish that this choice matters in general; we measure, for a user \emph{simulator}, whether the prompt or the weights reduces the gap to real users, and by how much.

User simulation for search ranges from rule-based interactive-IR toolkits to LLM agents that issue queries, click, give feedback, and ask clarifying questions~\cite{Zhang:2024:SIGIR,Zhang:2025:SIGIR,Owoicho:2023:SIGIR,Sekulic:2024:SIGIR-AP,Sekulic:2022:WSDM}, and to dedicated trained user language models~\cite{Naous:2025:arXiv,Wu:2026:arXiv}. Discriminator and utterance analyses of conversational-search simulators find that instruction-tuned assistant models make poor users, being too cooperative, wordy, and polite~\cite{Wang:2024:WWW,Sekulic:2024:TIST}. Trained user language models already outperform prompted ones for realism~\cite{Naous:2025:arXiv}. We add two results: a measured limit on prompt-level self-improvement, where a single hand-written prompt matches an iterative loop, and the effect on evaluation of removing that limit with weight updates. Our base follows a hybrid orchestration simulator~\cite{Zerhoudi:2025:CIKM}; the loop, the discriminator, and the validity results do not depend on that choice.

The closest prior work is DIAL~\cite{Zhu:2025:arXiv}, which improves a user simulator by adversarial preference optimization against a realism discriminator in mental-health support dialogue, driving the discriminator toward $0.5$ while restoring the diversity that supervised fine-tuning removes. Our weight-level variant reproduces that effect in conversational search, using rejection-sampling SFT rather than DIAL's DPO objective. Our addition is the contrast with prompt-level self-improvement and the effect on retrieval-effectiveness estimates; DIAL reports a related bias in dialogue-outcome rates, but tests neither a prompt baseline nor a graded retrieval metric.

Validating simulators rests on three modes, system-ranking agreement, behavioral-distribution match, and human realism~\cite{Bernard:2026:HAL,Zerhoudi:2022:ECIR}, with the training/evaluation objective divergence formalized separately~\cite{Balog:2025:SIGIR}. Profile-conditioned simulators reach ranking correlation $\rho\approx0.70$ with human judgments~\cite{Dou:2025:EMNLP}, though that conditioning is a fixed per-run profile that does not adapt during the interaction, and others measure the gap without closing it~\cite{Zhou:2026:arXiv}. A recent evaluation toolkit reports that the discriminator human-likeness check has little pooled predictive power for system-ranking validity~\cite{Zerhoudi:2026:SIGIR}, consistent with our finding that ranking validity is a property of the judge, so we use the discriminator to certify realism, not to predict ranking. Prior work establishes these properties separately; \textsc{AutoSim} combines behavioral-gap closing and ranking validity in one system on real conversational-search data, and places the prompt-versus-weight analysis beside them. Because part of \textsc{AutoSim} scores relevance with an LLM, we keep judge and generator in different model families, mindful of the known style, verbosity, and self-preference biases~\cite{Zheng:2023:NeurIPS,Panickssery:2024:NeurIPS,Saito:2023:CoRR}.

\section{The Sim-to-Real Gap}\label{sec:diagnosis}
%MG for me it was not clear in the next sentence what the simulator here is, so i had to reconstruct it from the resutls.
We first measure how the default simulator differs from real users on TREC iKAT\cite{Aliannejadi:2024:TREC,Aliannejadi:2025:TREC}. The measurements separate into two parts: its relevance judging already agrees with NIST, while its user behavior is far from real.

\subsubsection*{The behavioral gap}
A strong LLM's persona-conditioned relevance assessments already rank a controlled pool of retrieval systems in agreement with NIST (Kendall's $\tau$ up to $0.91$), and the agreement survives removing the persona, so ranking validity is not a property self-improvement has to create; \S\ref{sec:rq1} establishes it in full. What a more realistic simulator can still change is how the simulated \emph{user} behaves, and there the default is far from a real searcher: asked to play a search user, the model writes long, polite, fully formed requests, at $37.5$ words per message against $12.8$ for real users, with a held-out classifier separating its utterances from real ones at AUC $0.88$ and its queries retrieving the passages a real user retrieves almost never (overlap@5 $0.01$). The departure is in what the user says and asks for, not in how the simulator judges relevance.

\subsubsection*{The limit of prompting}
Editing the generation prompt moves the summary statistics while the underlying behavior remains unchanged: every variant we tried, a single hand-written terse instruction, an iterative reflective loop, and few-shot prompting with real terse utterances, still leaves a held-out classifier separating real from simulated sessions at AUC $\ge0.85$ (few-shot $0.94$, the default's own value) compared to a expected random baseline result of $0.5$, so the fixable gap is behavioral and lies beyond the prompt. %MG i reworked the sentence to make the setup clearer. Did you cite your own paper where you introduced this kind of evaluation the first time?

Together these measurements point to a single design: leave the relevance judge unchanged, target the utterance generator, and edit it at the weights rather than the prompt. \S\ref{sec:autosim} builds that simulator; \S\ref{sec:eval} confirms each step: the judge keeps the ranking valid (RQ1), weight-level editing closes the gap the prompt could not (RQ2), and only that closure removes an evaluation bias (RQ3).

\section{AutoSim}\label{sec:autosim}
\subsection{Simulator and editable components}
A prompted LLM simulator takes a user state and emits the next utterance under a fixed instruction. \textsc{AutoSim} keeps that generator and wraps it in state $S_t=\{P,I_t,H_t,C_{s,t}\}$: persona $P$ (a personal text knowledge base of background statements), information need $I_t$, dialogue history $H_t$, and a latent cognitive or affective state $C_{s,t}$ (satisfaction, frustration, patience). An orchestration policy $\pi_O\colon S_t\to C_k$ selects, at each turn, which component acts: deterministic simulators (query reformulation, click, stopping) for grounded routine behavior, or LLM cognitive functions (answer a clarifying question, give feedback, express satisfaction, generate the next utterance). We instantiate the base following a hybrid orchestration simulator~\cite{Zerhoudi:2025:CIKM}; since the components we edit exist in any LLM-based simulator, our results are not specific to this base.

This paper studies \emph{where} self-improvement should act, so we expose two components the loop can edit and hold everything else fixed. The first is the \emph{relevance-judging instruction}, the prompt the assessor applies; it sets the simulator's satisfaction and thus the ranking it induces. The second is the \emph{utterance generator}, the model that writes what the simulated user says. The generator can be edited in two ways: at the prompt, which changes its instruction, or at the weights, which changes its parameters. This prompt-versus-weights distinction is the variable our results depend on.

\subsection{Gap metric (CSG)}
We measure the simulator's gap to real users along two of the validity modes of~\cite{Bernard:2026:HAL}: system-ranking agreement and behavioral-distribution match. We call the combined measurement the Conversational-search Sim2Real Gap (CSG), which lets us test whether improving one mode also improves the other. A ranking term $R$ measures evaluation validity (\S\ref{sec:validity}), while behavioral terms measure realism. We measure the first behavioral dimension, communication style (D1, verbosity and politeness), in two ways, and the difference between them is itself a result. The first is a set of aggregate style statistics (mean words per message, politeness rate, short-message rate), which are easy to read off and easy to write into a prompt. The second is a held-out \emph{real-versus-simulated discriminator}, a classifier trained to tell a simulated utterance from a real one, reported as ROC AUC with $0.5$ meaning the two are indistinguishable; this is the adversarial realism criterion from generative-adversarial training~\cite{Goodfellow:2014:NIPS}, and unlike the summary statistics it cannot be met by matching a few numbers. We measure D1 in full, and we examine reformulation and clarification behavior (D2--D3) through query-need alignment, the overlap of the passages the simulator retrieves, and its mix of conversational moves (\S\ref{sec:substrate}); click, satisfaction, and stopping (D4--D6) are left to future work.

\subsection{Self-improvement Loop}
In each cycle, a controller records structured traces (every turn's policy choice, retrieval, click, and utterance), clusters the differences between simulated and real behavior, derives per-cluster evaluators with the judge drawn from a different model family than the generator, improves the chosen component by reflective population search, accepts an edit only when it does not regress on a calibration split, and monitors a rolling held-out set. The acceptance check is the loop's guarantee: it never keeps an edit that worsens calibration.

\subsection{Validity Testbed}\label{sec:validity}
We test whether \textsc{AutoSim} ranks IR systems as humans do on three independent TREC iKAT collections~\cite{Aliannejadi:2024:TREC,Aliannejadi:2025:TREC}: iKAT 2023 (176 judged turns), iKAT 2024 (116 turns), and the multi-session iKAT 2025 (45 turns), each with NIST passage qrels over the same ClueWeb22-B corpus~\cite{Overwijk:2022:SIGIR}. iKAT is the only public collection with all four of the properties we need at once: multi-turn conversational structure, per-user PTKB personas, passage-level NIST qrels, and a corpus we can query at run time. On each year we build a \emph{controlled pool of twelve constructed systems} (retrieval pipelines we assemble) over ClueWeb22-B, spanning a wide effectiveness range: first stages of BM25, learned-sparse SPLADE, and reciprocal-rank fusion; over raw, context-resolved, and PTKB-expanded queries; with and without a cross-encoder reranker. For iKAT 2023, where two organizer runs (a BM25+RM3 pipeline with automatic and with manual rewriting) are available with pooled judgments, we fold them in for a fourteen-system pool with real participant systems.

The reference ranking orders systems by mean condensed nDCG@5 over the NIST qrels, condensed so that non-pooled systems are not penalized for retrieving relevant but unjudged passages~\cite{Bernard:2026:HAL}; we report raw nDCG@5 alongside. \textsc{AutoSim}'s induced ranking orders systems by its persona-conditioned mean satisfaction with each system's top-5 passages. We report Kendall's $\tau$ (ties-corrected) with a Monte-Carlo permutation test and a system-level bootstrap CI that resamples the systems, since the system is the unit of the test. SPLADE and BM25 run over a $116$M-passage ClueWeb22-B index, reranking and assessment use production LLMs, and the generator and judge come from different model families.

\subsection{Setup and reproducibility}
The loop runs reflective population search: each cycle proposes $N$ candidate edits to the chosen component (generated by the judge from the observed gap), scores each on a calibration split, selects the best, and keeps it only if it does not regress. We validate on a disjoint held-out split fixed before any run, and compare against \textsc{static} (no self-improvement) and \textsc{no-calibration} (edits accepted without the calibration check). A strong model's default style is verbose, so we initialize the generator prompt from a deliberately weak style, and report both the gain from that start and the distance from the model's default. To edit the generator at the weights, we warm-start a small open generator (Qwen2.5-3B with a LoRA adapter) on real iKAT utterances, then run adversarial rejection-sampling rounds: each round samples several utterances per held-out turn, keeps the ones a frozen real-versus-simulated discriminator rates most human, fine-tunes on that set, refreshes the discriminator, and repeats for four rounds~\cite{Zhu:2025:arXiv}. Freezing the discriminator within a round and taking the previous round as the reference keeps the target from moving.

Held-out utterances are the $82$ iKAT turns disjoint from the SFT split (seed-fixed); the discriminator sees $2$ samples per held-out turn against the matched real utterances, and every AUC is scored five-fold with folds grouped by turn, so a turn's real and simulated utterances never split across train and test. The weight-level generator is Qwen2.5-3B/7B-Instruct (and, for the cross-family check, Mistral-7B-Instruct-v0.3) with a LoRA adapter ($r{=}16$, $\alpha{=}32$, $q$/$v$ projections), warm-started and then improved over four adversarial rounds. The in-loop discriminator is a logistic classifier over multilingual-MiniLM embeddings; the second discriminator is a fine-tuned XLM-RoBERTa-base. The validity pipeline uses two open-weight instruction models from different families (an $80$B-class Qwen generator and a $31$B-class Gemma judge), both at temperature $0$ and served without any proprietary API, so the whole pipeline runs on released weights. The RQ1 rank-correlation, RQ2 discriminator (Table~\ref{tab:substrate}), and RQ3 effectiveness-bias numbers are all reproducible from committed result files, and we release the configurations and the trained LoRA adapters~\footref{fn:repo}.

\section{Evaluation}\label{sec:eval}
The evaluation answers three questions, each building on the last. \textbf{RQ1}: does \textsc{AutoSim} rank IR systems as NIST does (\S\ref{sec:rq1})? \textbf{RQ2}: can self-improvement close the behavioral gap to real users, and at the prompt or the weights (\S\ref{sec:substrate})? \textbf{RQ3}: does closing the gap make the simulator a better evaluator (\S\ref{sec:bias})?

\subsection{RQ1: Is \textsc{AutoSim} a valid evaluator?}\label{sec:rq1}
\textsc{AutoSim}'s persona-conditioned satisfaction agrees with the NIST ranking at $\tau{=}0.66$ on the twelve-system pool (condensed nDCG@5; $\tau{=}0.50$ on raw), Monte-Carlo permutation $p{=}0.002$. The system-level bootstrap 95\% CI on $\tau$ is $[0.30,0.90]$, wide as a twelve-system test must be and excluding zero; no single system drives the ranking (leave-one-out $\tau\in[0.59,0.73]$), and re-scoring the pool five times gives $\tau{=}0.72{\pm}0.01$ (Table~\ref{tab:robust}). Adding the two real participant runs (Table~\ref{tab:validity}, starred) leaves the picture much the same: $\tau{=}0.65$ over fourteen systems ($p{=}0.001$, leave-one-out $\tau\in[0.59,0.77]$), $\tau$ stays at $0.59$ even when the weak \texttt{auto} run is removed, and \textsc{AutoSim} ranks that weak official run last exactly as NIST does. The agreement is aggregate and not perfect for every pair: \textsc{AutoSim} under-ranks the strong \texttt{manual} run, placing it 10th of 14 where NIST places it 3rd, its largest single disagreement. As a sanity check, naive PTKB query-stuffing degrades retrieval, so the \texttt{(ptkb)} rows fall in the reference ranking, and \textsc{AutoSim} independently ranks them low.

\begin{table}[t]\centering\small
\caption{RQ1. Fourteen real iKAT 2023 systems, ranked by \textsc{AutoSim} against NIST.}
\label{tab:validity}
\begin{threeparttable}
\begin{tabular*}{\columnwidth}{@{\extracolsep{\fill}}lrrr}
\toprule
System & nDCG@5 & cond.\ nDCG & AutoSim \\
\midrule
\rowcolor{gray!15}\multicolumn{4}{@{}l}{\textbf{Context-resolved queries}\;\footnotesize(best retrieval)}\\% lint-ok
rrf(resolved)            & 0.287 & 0.343 & 0.632 \\
splade(resolved)         & 0.210 & 0.295 & 0.661 \\
bm25(resolved)           & 0.225 & 0.276 & 0.553 \\
rrf{+}rerank             & 0.124 & 0.228 & 0.594 \\
bm25{+}rerank            & 0.107 & 0.200 & 0.506 \\
splade{+}rerank          & 0.093 & 0.199 & 0.604 \\
\addlinespace
\rowcolor{gray!15}\multicolumn{4}{@{}l}{\textbf{PTKB-expanded queries}\;\footnotesize(query stuffing)}\\% lint-ok
splade(ptkb)             & 0.090 & 0.148 & 0.553 \\
rrf(ptkb)                & 0.066 & 0.133 & 0.506 \\
splade(ptkb){+}rerank    & 0.050 & 0.105 & 0.546 \\
bm25(ptkb)               & 0.035 & 0.058 & 0.289 \\
\addlinespace
\rowcolor{gray!15}\multicolumn{4}{@{}l}{\textbf{Raw queries}\;\footnotesize(no context)}\\% lint-ok
splade(raw)              & 0.062 & 0.096 & 0.334 \\
bm25(raw)                & 0.088 & 0.095 & 0.277 \\
\addlinespace
\rowcolor{gray!15}\multicolumn{4}{@{}l}{\textbf{Real organizer runs}}\\% lint-ok
bm25\_rm3(manual)$^\star$ & 0.289 & 0.289 & 0.505 \\
bm25\_rm3(auto)$^\star$   & 0.033 & 0.033 & 0.155 \\
\bottomrule
\end{tabular*}
\begin{tablenotes}[flushleft]\footnotesize
\item Twelve constructed systems grouped by query type, plus two organizer runs ($^\star$); $176$ turns. \emph{Resolved} rewrites the turn to stand on its own by filling in the conversational context; \emph{PTKB} stuffs the query with persona statements; \emph{raw} uses the turn as-is. $\tau$(\textsc{AutoSim}, condensed nDCG@5)$=0.648$, $p{=}0.001$; sorted by condensed nDCG within each group.
\end{tablenotes}
\end{threeparttable}
\end{table}

Re-running the same twelve-system protocol and live ClueWeb22-B pipeline on two further collections shows the result holds beyond 2023: on iKAT 2024 (116 turns) \textsc{AutoSim} agrees more strongly ($\tau{=}0.91$ condensed, $0.85$ raw; $p{<}0.001$; CI $[0.73,1.00]$; leave-one-out $\tau\in[0.89,0.93]$), and on the multi-session iKAT 2025 (45 turns) it again reaches $\tau{=}0.91$ condensed ($0.76$ raw, $p{<}0.001$, CI $[0.73,1.00]$). Where validity is weakest is year-dependent: restricted to the six closest-quality systems, $\tau$ falls to $0.33$ on 2023 ($p{=}0.47$, n.s.) but recovers to $0.87$ on 2024 ($p{=}0.02$, suggestive rather than family-wise corrected), so near-tie resolution, not the ranking itself, is what varies. Off ClueWeb22-B, on TREC CAsT 2020~\cite{Dalton:2020:TREC} over MS MARCO~\cite{Nguyen:2016:CoRR}, run persona-free over a ten-system pool, satisfaction still tracks NIST ($\tau{=}0.56$, $p{=}0.03$, CI $[0.11,0.85]$), and it holds when a neural cross-encoder rerank tier joins the pool ($\tau{=}0.58$ over fifteen systems, $p{=}0.002$, CI $[0.28,0.79]$; even the five neural systems on their own rank in the right order at $\tau{=}0.80$), so the agreement does not depend on a lexical-only pool; the signal is weaker than on iKAT and limited by about half corpus coverage, but positive on a corpus \textsc{AutoSim} was not built on.

A persona-free relevance judge over the same passages ranks the pool at $\tau{=}0.79$, on par with the persona-conditioned $0.72$ (both the $T{=}0$ re-scored mean; Table~\ref{tab:robust}), so the validity does not depend on the persona; re-scoring the 2024 pool with a different-family assessor (Gemma against the Qwen default) reproduces the ranking unchanged ($\tau{=}0.91$), the two assessors agreeing at $\tau{=}0.94$. The agreement comes from relevance-assessment ability, which strong LLMs already have; the persona is what makes \textsc{AutoSim} a \emph{user} and carries its behavioral realism. Across the collections we tried, \textsc{AutoSim} ranks systems as NIST does wherever they differ clearly in quality, an agreement its relevance judgments drive.

\begin{table}[t]\centering\small
\caption{RQ1. Validity is robust across three iKAT years and a second corpus.}
\label{tab:robust}
\begin{threeparttable}
\begin{tabular*}{\columnwidth}{@{\extracolsep{\fill}}lrl}
\toprule
Setting & $\tau$ & note \\
\midrule
\rowcolor{gray!15}\multicolumn{3}{l}{\textbf{iKAT 2023 (176 turns)}}\\% lint-ok
\;12 constructed systems          & \underline{\textbf{0.66}} & perm.\ $p{=}0.002$, CI $[0.30,0.90]$ \\
\;\;+2 real participant runs (14)  & 0.65 & $p{=}0.001$, CI $[0.26,0.91]$ \\
\;re-scored $\times5$ (T$=$0)       & 0.72 & $\pm0.01$ (reproducible) \\
\;persona-free relevance judge     & 0.79 & persona not the source of validity \\
\;6 closest-quality systems        & 0.33 & $p{=}0.47$ (n.s.): near-ties hard \\
\midrule
\rowcolor{gray!15}\multicolumn{3}{l}{\textbf{iKAT 2024 (116 turns), same protocol}}\\% lint-ok
\;12 constructed systems          & \underline{\textbf{0.91}} & perm.\ $p{<}0.001$, CI $[0.73,1.00]$ \\
\;6 closest-quality systems        & 0.87 & $p{=}0.02$: near-ties resolved \\
\midrule
\rowcolor{gray!15}\multicolumn{3}{l}{\textbf{iKAT 2025 (45 turns), multi-session}}\\% lint-ok
\;12 constructed systems          & \underline{\textbf{0.91}} & perm.\ $p{<}0.001$, CI $[0.73,1.00]$ \\
\midrule
\rowcolor{gray!15}\multicolumn{3}{l}{\textbf{cross-corpus: CAsT 2020 / MS MARCO (10 sys)}}\\% lint-ok
\;different corpus, no personas    & \underline{\textbf{0.56}} & $p{=}0.03$, CI $[0.11,0.85]$ \\
\bottomrule
\end{tabular*}
\begin{tablenotes}[flushleft]\footnotesize
\item Kendall-$\tau$ on condensed nDCG@5 under stress; CI is the system-level bootstrap on $\tau$. The persona-free row shows validity does not come from the persona.
\end{tablenotes}
\end{threeparttable}
\end{table}

\begin{figure}[t]\centering
\includegraphics[width=0.85\columnwidth]{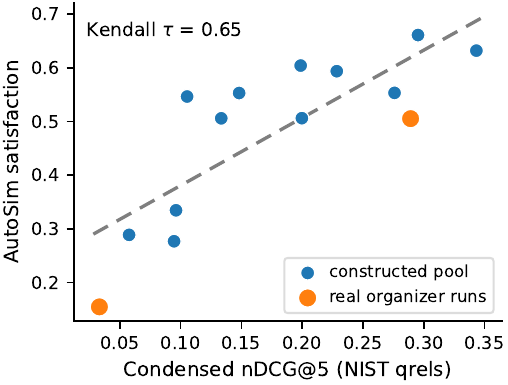}
\caption{RQ1. Sim-to-real validity: \textsc{AutoSim}'s satisfaction tracks NIST condensed nDCG@5 across fourteen systems (Kendall $\tau{=}0.65$, $p{=}0.001$). The two orange points are the real iKAT organizer runs (\texttt{bm25+rm3} over manually vs.\ automatically rewritten queries); \textsc{AutoSim} rates the weak automatic run lowest, as NIST does.}\label{fig:validity}
\end{figure}

\subsection{RQ2: Can self-improvement close the behavioral gap, at the prompt or the weights?}\label{sec:substrate}
The diagnosis (\S\ref{sec:diagnosis}) placed the fixable gap in the utterance generator and showed the prompt does not close it. We now hold the loop to that gap and measure how far each, prompt-level and weight-level, gets on the two measures of \S\ref{sec:autosim}.

At the prompt, across twelve seeds, the loop raises held-out communication-style fidelity above the no-evolution baseline: final-cycle CSG is $0.76{\pm}0.10$ for the full loop against $0.56{\pm}0.07$ for \textsc{static}, a gain significant under a paired $t$-test ($p{=}0.0002$). Table~\ref{tab:style} shows what it fixes: the default user averages $37.5$ words against $12.8$ for real users, self-improvement brings it to $11.0$, and the length distribution moves much closer to human (Kolmogorov--Smirnov $D{=}0.24$ against $0.81$). Two limits qualify this. The blind \textsc{no-calibration} variant reaches only $0.63{\pm}0.07$ and does not clear \textsc{static} at significance ($p{=}0.08$), so the calibration signal is what makes the loop reliably improve, and the acceptance check adds a guarantee against regression. And style is not matched exactly: the tuned style overshoots politeness ($39\%$ against $12\%$) and misses the $5\%$ of very-short human messages, so its distribution stays distinguishable ($p{=}0.022$). The loop closes the gap that dominates and leaves the rest.

A sharper comparison is against a single hand-written prompt. On the aggregate statistics the loop optimizes, the loop does worse: a one-line prompt (``write a short message, about $12$ words, and skip pleasantries'') reaches CSG-D1 $0.80$ against the loop's $0.68$ on the same turns, because three summary statistics are easier to write directly than to reach through search. On the discriminator the loop does better (AUC $0.85$ against above $0.9$ for the default and terse prompt, on both an embedding and a lexical classifier), but still does not reach $0.5$. Few-shot prompting with real terse utterances, the strongest prompt baseline, matches real length ($12.3$ words) yet a held-out classifier still separates it at AUC $0.94$, the default's value.

\begin{table}[t]\centering\small
\caption{RQ2. Self-improvement depends on where it acts (iKAT, held-out).}\label{tab:substrate}
\begin{threeparttable}
\begin{tabularx}{\columnwidth}{Lc}
\toprule
self-improvement & discriminator AUC \\
\midrule
none (default prompt) & $0.88$ \\
prompt-level (best of four methods) & $0.84$ \\
\rowcolor{gray!15}weight-level, warm SFT on real & $\mathbf{0.45}$ \\% lint-ok
\rowcolor{gray!15}weight-level, {+}\,adversarial rounds & $0.42$--$0.50$ \\% lint-ok
\bottomrule
\end{tabularx}
\begin{tablenotes}[flushleft]\footnotesize
\item Real-versus-simulated discriminator AUC on the same base generator; $0.5$ is indistinguishable from real, higher is more separable. A held-out XLM-RoBERTa discriminator and two further base models confirm both extremes (\S\ref{sec:substrate}).
\end{tablenotes}
\end{threeparttable}
\end{table}

\begin{table}[t]\centering\small
\caption{RQ2. Self-improvement improves the utterance generator but finds no accepted edit for the relevance judge.}
\label{tab:selfimprove}
\begin{threeparttable}
\begin{tabularx}{\columnwidth}{Lr}
\toprule
Component / variant & held-out fidelity \\
\midrule
\rowcolor{gray!15}\textbf{Behavioral, full (calibration check)}     & $\mathbf{0.76}\pm0.10$ \\% lint-ok
Behavioral, no-calibration   & $0.63\pm0.07$ \\
Behavioral, static           & $0.56\pm0.07$ \\
Assessor, full (calibration check, $\tau$) & $0.61$ (no surviving edit) \\
\bottomrule
\end{tabularx}
\begin{tablenotes}[flushleft]\footnotesize
\item Behavioral: held-out CSG-D1 final cycle, mean$\pm$SD over 12 seeds; the full loop improves on \textsc{static} at $p{=}0.0002$ (paired $t$), the blind variant does not ($p{=}0.08$). Assessor: held-out ranking $\tau$; the check finds no surviving edit.
\end{tablenotes}
\end{threeparttable}
\end{table}

Moving the same objective to the weights, the discriminator can no longer separate the weight-level simulator from real users: AUC falls from $0.88$ to $0.45$ and holds there across four adversarial rounds (Table~\ref{tab:substrate}). The utterances behind the drop stay diverse: they are as mutually distinct as real ones (mean pairwise token overlap $0.043$ against $0.044$, every utterance unique), match their length variety (entropy $3.1$ against $3.1$) and mean length ($12.3$ against $13.3$ words), and only $1.2\%$ copy a training utterance, though their vocabulary is somewhat narrower (word distinct-2 $0.54$ against $0.84$). The match extends beyond phrasing to three further behavioral measures. In \emph{what the user retrieves}, the weight-level queries retrieve the same ClueWeb passages as the real query an order of magnitude more often than any prompt-level variant (overlap@5 $0.17$ against $0.01$ default and $0.015$ terse). In \emph{how the user moves}, classified into new request, refinement, feedback, and clarification, the weight-level move mix matches real users (total-variation distance $0.07$) where the default diverges sharply ($0.34$--$0.46$ across two classifier families), taking the clarifying and feedback turns of an assistant in place of the focused refinement of a searcher; both classifiers put the weight-level gap at $0.07$, though per-utterance labels agree only weakly ($\kappa{=}0.22$), so we read the move mix distributionally. In \emph{how the user reformulates}, we measure query-need alignment as the word overlap between a variant's next-turn query and its resolved full-need form: real users write need-focused queries (overlap $0.56$) and the weight-level simulator matches them ($0.51$), while the verbose default dilutes the need with conversational padding ($0.23$) and even a deliberately terse prompt, the shortest variant at $9.6$ words yet the least aligned, does not recover it ($0.19$), so a short query is not by itself need-focused.

The weight-level result raises two objections, which we check directly. \emph{Is the discriminator too weak?} With folds grouped by turn, a fine-tuned XLM-RoBERTa separates the default almost perfectly (AUC $0.94$) yet cannot separate the weight-level one ($0.51{\pm}0.06$), and the same discriminator tells a lightly politeness-perturbed real utterance from an unperturbed one at AUC $0.99$, so it is strong, and the $0.51$ shows the two are indistinguishable. \emph{Is the result specific to one model?} The whole procedure on a Qwen2.5-7B base reproduces both extremes ($1.00$ against $0.50$), and on an independent Mistral-7B base again ($0.97$ against $0.40$). The indistinguishability even holds against strong judges: two frontier LLMs asked to pick the human beat chance on the default ($0.59$, $0.89$) but not on the weight-level one ($0.50$, $0.48$, n.s.), a panel reading the full conversation stays near $0.5$ on it ($0.55$, $p{=}0.44$) while flagging the default ($0.75$), and three human annotators catch the default on $28$--$30$ of $30$ trials but only $4$--$11$ of $30$ for the weight-level one, all below the $15/30$ chance level, so even a human reading the whole dialogue cannot separate it from a real user. The learned style also transfers: applying the 2023-tuned style to iKAT 2024 brings the default's $38.6$ words down to $11.8$ against $11.7$ for real users, indistinguishable in mean length (permutation $p{=}0.89$ against $p{<}10^{-4}$ for the default).

Self-improvement therefore reaches a limit at the prompt but not at the weights: holding the base generator fixed, prompt-level self-improvement does not lower the discriminator below AUC $0.85$, whereas weight-level self-improvement on that same generator lowers it to $0.5$. The relevance judge shows a third pattern: editing the assessor prompt under the check produces no accepted edit (Table~\ref{tab:selfimprove}), and unchecked edits move $\tau$ by up to about $0.1$, which is within the sampling noise, since a strong model is already a competent judge (\S\ref{sec:rq1}). Self-improvement thus has room to change how the user behaves but not how it judges; ranking validity comes from a strong judge, and self-improvement is best applied to the generator. This is consistent with the training and evaluation objectives not improving together~\cite{Balog:2025:SIGIR}, though it is not the same statement, and it does not imply that judges cannot be improved: methods that retrain the judge~\cite{Wang:2024:arXiv} act on the component our prompt-level check leaves unchanged.

\begin{table}[t]\centering\small
\caption{RQ2. What the behavioral loop fixes on communication style.}
\label{tab:style}
\begin{threeparttable}
\begin{tabularx}{\columnwidth}{Lrrr}
\toprule
 & avg.\ words & polite\,\% & very-short\,\% \\
\midrule
Real users (human)        & 12.8 \,[11.3,14.3] & 12 & 5 \\
\textsc{AutoSim} default  & 37.5 \,[35.2,39.7] & 21 & 0 \\
\rowcolor{gray!15}\textbf{\textsc{AutoSim} tuned}  & \textbf{11.0} \,[10.2,11.7] & 39 & 0 \\% lint-ok
\bottomrule
\end{tabularx}
\begin{tablenotes}[flushleft]\footnotesize
\item Message-style features against real iKAT users; $n{=}80$ of the held-out turns, brackets $=$ bootstrap 95\% CI on the mean.
\end{tablenotes}
\end{threeparttable}
\end{table}

\begin{figure}[t]\centering
\includegraphics[width=0.82\columnwidth]{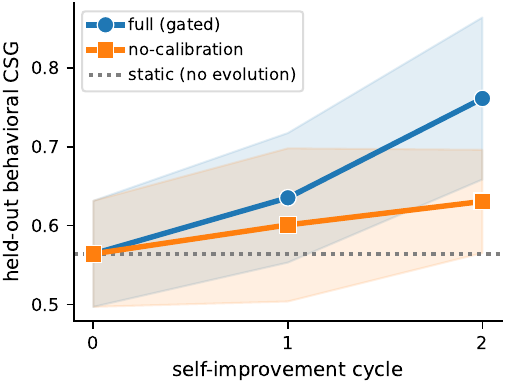}
\caption{RQ2. Behavioral self-improvement across 12 seeds (mean$\pm$SD band): the checked loop rises above \textsc{static} (paired $t$, $p{=}0.0002$); the blind loop's gain is not significant ($p{=}0.08$).}\label{fig:selfimprove}
\end{figure}

\subsection{RQ3: Does closing the gap make a better evaluator?}\label{sec:bias}
Closing the behavioral gap matters only if it makes the simulator a better evaluator. System-ranking $\tau$ does not move: the ranking is already valid (\S\ref{sec:rq1}) and does not depend on which simulator writes the query. The effect appears instead on a quantity the weight-level training never optimized, the absolute retrieval effectiveness a simulator's own queries imply. Effectiveness comes from running retrieval on the same utterances, so the two are related. The training targets distributional similarity rather than nDCG, and per-turn effectiveness correlates only weakly with the real user's ($r{=}0.44$), so the match is distributional rather than mechanical. On $157$ held-out judged turns we run BM25 retrieval with each simulator's query and compare its nDCG@5 to the real user's. The bias is largest on the harder, fully held-out 2024 collection, where real users' own queries retrieve little (mean nDCG $0.016$): a default, wordy simulator, whose retrieval query averages $28$ words, over-estimates effectiveness by $+0.136$ ($95\%$~CI $[0.10,0.17]$, $p{<}10^{-3}$, Table~\ref{tab:bias}).

The simplest explanation is length, since real users are terse, and it does not account for the effect. A hand-written terse prompt ($9$ words) still over-estimates by $+0.087$ and a length-matched truncation of the verbose query by $+0.071$, both with confidence intervals above zero, while only the weight-level simulator is not detectably biased ($+0.014$, $[{-}0.00,0.03]$, $p{=}0.41$); across query sources length explains just $2\%$ of the variance in retrieved nDCG ($r^2{=}0.02$). Matching verbosity is not enough; matching the full query distribution removes the bias. On the easier 2023 held-out subset, where real users retrieve reasonably (mean $0.080$), no simulator is detectably biased (default $+0.009$, n.s.), so the over-estimate appears exactly where real users struggle and an evaluation most needs to be trusted; pooled over both collections the default over-estimates by $+0.103$.

This over-estimate is not a harmless level shift that could be calibrated away, because it hides the failures an evaluation exists to reveal. Real users retrieve nothing relevant on $86\%$ of these turns; on exactly those turns a default simulator's query still returns a relevant passage $57\%$ of the time, while the weight-level simulator returns one only $14\%$ of the time and fails where real users fail. The result is a wrong decision: on 2024 an evaluation run with a default simulator reads the system's effectiveness as $0.15$ nDCG, well above a modest $0.05$ deployment threshold, where real users get $0.02$ and the weight-level simulator reports the same failure ($0.03$). A realistic simulator, built at the weight level, gives a more accurate estimate of how a system serves real users, the quantity an aggregate rank correlation hides.

\subsection{Limitations and threats to validity}
Our pool is a constructed one plus two real organizer runs, short of the full participant field, and the behavioral reference is organizer-authored iKAT utterances, a proxy for logged end-users; the human evidence rests on three annotators (\S\ref{sec:substrate}), who agree with the discriminators ($28$--$30$ of $30$ on the default, $4$--$11$ of $30$ on the weight-level one); we release the protocol for a larger study. The validity holds for systems that differ clearly in quality: near-tie resolution is reliable on 2024 but only marginal on 2023, whose $\tau$ CI is wide, and the 2025 result rests on 45 judged turns; we read the borderline tests in Table~\ref{tab:robust} ($p{=}0.02$--$0.03$) as descriptive rather than family-wise corrected. The behavioral results cover four axes across D1--D3, leaving click, satisfaction, and stopping (D4--D6) for future work, and the cross-corpus step (CAsT 2020, $\tau{=}0.56$--$0.58$, $p{\le}0.03$) is limited by about half corpus coverage. We control for contamination with disjoint calibration and validation splits fixed in advance, and apply weight-level self-improvement only to generation; the relevance judge, near its ceiling under our prompt-level check, may still move under methods that retrain it.

\begin{table}[t]\centering\small
\caption{RQ3. Evaluation bias by collection (iKAT held-out).}\label{tab:bias}
\begin{threeparttable}
\begin{tabularx}{\columnwidth}{Lccc}
\toprule
query source & words & iKAT 2024\tnote{a} & 2023 h.o.\tnote{b} \\
\midrule
default (verbose) & $28$ & $+0.136$ $[0.10,0.17]$   & $+0.009$ \\
terse prompt      & $9$  & $+0.087$ $[0.06,0.12]$   & ${-}0.028$ \\
length-matched    & $11$ & $+0.071$                 & ${-}0.042$ \\
\rowcolor{gray!15}\textbf{weight-level} & $12$ & $\mathbf{+0.014}$ $[{-}0.00,0.03]$ & $+0.015$ \\% lint-ok
\bottomrule
\end{tabularx}
\begin{tablenotes}[flushleft]\footnotesize
\item Bias $=$ mean nDCG@5 of a simulator's query minus the real user's; $0$ unbiased, $+$ over-estimates. Brackets: bootstrap 95\% CI; significance by paired Wilcoxon; a CI excluding $0$ is a detectable bias.
\item[a] iKAT 2024, $n{=}116$ (real mean nDCG $0.016$).
\item[b] 2023 held-out, $n{=}41$ (real mean $0.080$); all values n.s.\ (CIs include $0$).
\end{tablenotes}
\end{threeparttable}
\end{table}

\section{Conclusion}
Editing the prompt cannot close a simulator's behavioral gap to real users; editing the weights can; and that closure removes an evaluation bias. We show this on \textsc{AutoSim}, a conversational-search user simulator that is first a valid IR evaluator ($\tau{=}0.66/0.91/0.91$ across three iKAT years, reproducible, stable with real participant runs, and holding across corpora and assessor families) whose validity comes from relevance-assessment competence rather than the persona. The limit depends on where self-improvement acts: prompt-level self-improvement, hand-written, tuned, or few-shot, plateaus at a real-versus-simulated discriminator AUC near $0.85$, while weight-level self-improvement drives it to $0.5$ across two model families and four behavioral axes. The closure matters because a default simulator over-estimates retrieval effectiveness by $+0.136$ nDCG where real users struggle and hides $57\%$ of their retrieval failures, while the weight-level simulator is not detectably biased; a realistic simulator, built at the weight level, is a more valid evaluator. Both effects are visible only on a learned discriminator and an effectiveness-bias test, a caution for anyone measuring simulator quality on summary metrics. Future work extends the behavioral axes to satisfaction and stopping, replaces the organizer-authored reference with logged users, and broadens the cross-corpus step to fuller corpus coverage. We release the framework, the protocol, and the models~\footref{fn:repo}.

% \begin{acks}

% \noindent
% This work has received funding from the Bavarian State Ministry of Economic Affairs, Regional Development, and Energy (StMWi) under the funding code VAL-2505-0003.

% \end{acks}

\bibliographystyle{ACM-Reference-Format}
\bibliography{reference}

\end{document}